\documentclass[12pt,a4paper,pdfusetitle]{article}
\usepackage[utf8]{inputenc}
\usepackage[T1]{fontenc}
\usepackage{jheplike,tocloft,tocloft,amsmath,amssymb,amsfonts,mathrsfs,mathtools,xcolor,graphicx}
\usepackage[backgroundcolor=black!5, linecolor=green!60]{todonotes}
\usepackage{booktabs}

\hypersetup{linkcolor=[rgb]{0.15,0.35,0.75},colorlinks=true,citecolor=[rgb]{0.675,0,0.2},urlcolor=[rgb]{0.15,0.35,0.65}}
\DeclareMathOperator{\Res}{Res}

\DeclareMathOperator{\PSL}{PSL}

\DeclareMathOperator{\Aut}{Aut}
\DeclareMathOperator{\Fix}{Fix}
\DeclareMathOperator{\id}{id}

\renewcommand{\P}[1]{\ensuremath{\mathbb{P}^{#1}}}

\newcommand{\fwbox}[2]{\text{\makebox[#1][c]{$\hspace{-150pt}\displaystyle#2\hspace{-150pt}$}}}
\newcommand{\fwboxL}[2]{\text{\makebox[#1][l]{$#2$}}}
\newcommand{\fwboxR}[2]{\text{\makebox[#1][r]{$#2$}}}
\newcommand{\equivR}{\fwbox{15.5pt}{\hspace{-0pt}\fwboxR{0pt}{\raisebox{0.47pt}{\hspace{1.25pt}:\hspace{-4pt}}}=\fwboxL{-0pt}{}}}

\title{Traintrack Calabi-Yaus from Twistor Geometry}
\author[a]{Cristian~Vergu,}\emailAdd{c.vergu@nbi.ku.dk}
\author[a]{Matthias~Volk}\emailAdd{mvolk@nbi.ku.dk}
\affiliation[a]{Niels Bohr International Academy and Discovery Center, Niels Bohr Institute,\\University of Copenhagen, Blegdamsvej 17, DK-2100, Copenhagen \O, Denmark}

\abstract{%
  We describe the geometry of the leading singularity locus of the traintrack integral family directly in momentum twistor space.
  For the two-loop case, known as the elliptic double box, the leading singularity locus is a genus one curve, which we obtain as an intersection of two quadrics in \(\mathbb{P}^{3}\).
  At three loops, we obtain a K3 surface which arises as a branched surface over two genus-one curves in \(\mathbb{P}^{1} \times \mathbb{P}^{1}\).  We present an analysis of its properties.
  We also discuss the geometry at higher loops and the supersymmetrization of the construction.
}
\preprint{}

\begin{document}
\maketitle

\section{Introduction}
\label{sec:introduction}

While it was initially hoped that the integrals which appear in computations in planar \(\mathcal{N} = 4\) SYM are expressible in terms of generalized polylogarithms, it has by now become clear that this is not the case.\footnote{Work on the Kontsevich conjecture by Belkale and Brosnan~\cite{belkale2003} had given good reasons to be pessimistic.  More recently, Brown and Schnetz~\cite{Brown:2010bw} have given explicit examples in \(\phi^4\) theory, which contain K3 geometries.}  Not only are the generalized polylogarithms insufficient but, by any reasonable measure, most of the integrals in \(\mathcal{N} = 4\) SYM seem to require more complicated classes of functions, which are as of yet very poorly understood.

One class of integrals which is relatively well-understood is the class of pure integrals.  These integrals have leading singularities (see ref.~\cite{Cachazo:2008vp}) which are pure numbers such as \(0\) or \(\pm 1\).  In all known examples they are computable in terms of generalized polylogarithms.

Recall that to obtain leading singularities one takes residues in the propagators of the integral.  Doing so, Jacobian factors are generated in which one can often take further residues.  If we start with an integral with fewer propagators than integration variables, two things can happen.  Either one can generate enough Jacobian factors to take residues in, so that the integral localizes, or not.  If the integral does not localize, then the process of taking residues ends with a holomorphic form.  This form may however develop poles for special kinematics.

\begin{figure}[t]
  \centering
  \includegraphics[width=0.9\textwidth]{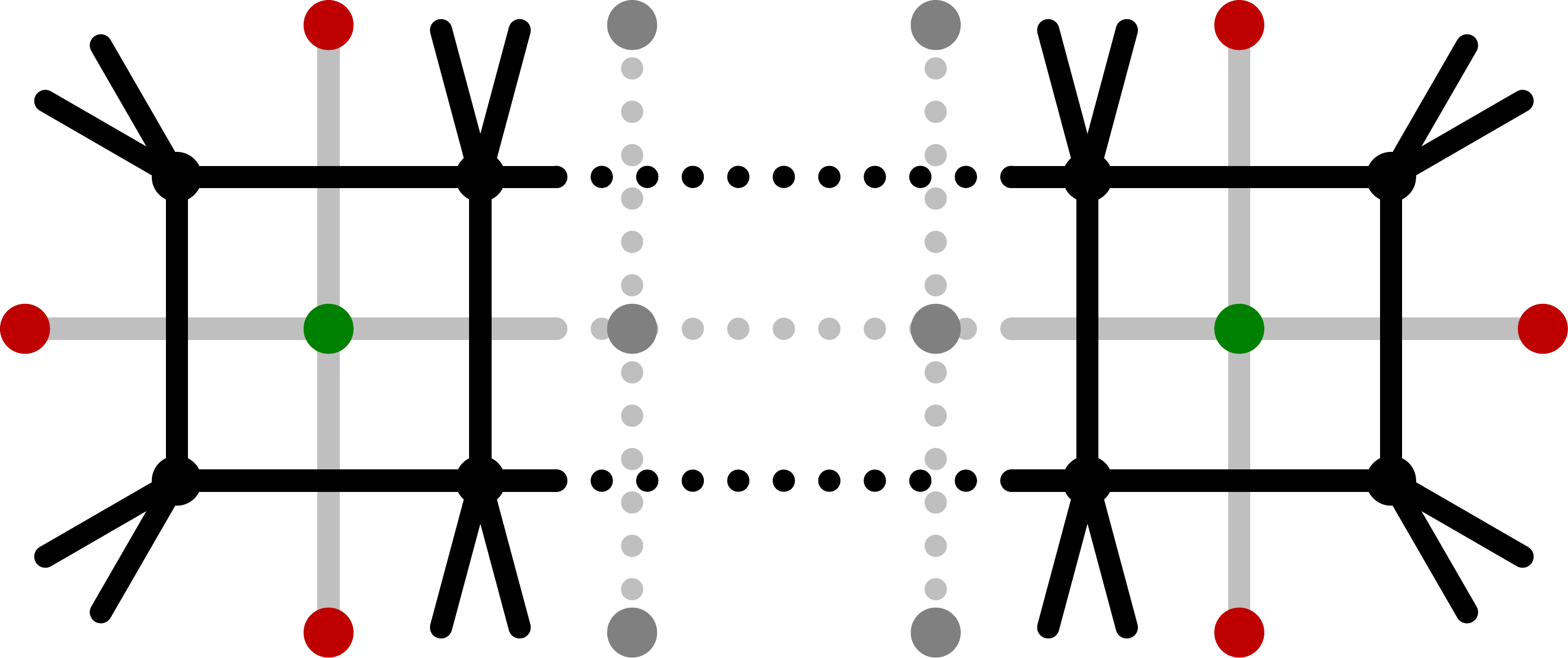}
  \label{fig:traintrack-integrals}
  \caption{The traintrack integrals}
\end{figure}

The leading singularity locus, when it is not a set of points, turns out to be an interesting variety of Calabi-Yau type.  The discussion above makes it plausible that one is more likely to find integrals which do not localize if we consider examples with as few propagators as possible.  Since triangles are not possible in a dual-conformal expansion in planar \(\mathcal{N} = 4\) SYM, the examples we consider are box integrals.  As it turns out, ladder integrals are computable in terms of classical polylogarithms (see ref.~\cite{Usyukina:1993ch}).  The simplest integral which can not be localized by taking residues is the elliptic double box integral, studied in refs.~\cite{CaronHuot:2012ab, Bourjaily:2017bsb}.  It is part of a family of integrals, called traintrack integrals (see fig.~\ref{fig:traintrack-integrals}).
There are many other examples in the literature, where Calabi-Yau geometries have been identified in loop integrals, see e.g.~\cite{Brown:2009ta,Brown:2010bw,Bourjaily:2018yfy,Besier:2019hqd,Festi:2018qip,Bourjaily:2019hmc,Klemm:2019dbm,Bloch:2014qca,Bloch:2016izu}.

The traintrack integrals were studied in ref.~\cite{Bourjaily:2018ycu}.  This reference studied three- and four-loop integrals using Feynman parametrization.  The leading singularity loci were defined as hypersurfaces in various weighted projective spaces, whose coordinates were related to the Feynman parameters of the original integral.  The constructions in ref.~\cite{Bourjaily:2018ycu} were pretty involved, in that they required complicated changes of variables which did not seem to fit a pattern that could be generalized to all loops.

In this paper we study the leading singularity locus by using the momentum twistor description of the traintrack integrals.  Momentum twistors were introduced by Hodges~\cite{Hodges:2009hk} in order to make the dual conformal symmetry~\cite{Drummond:2006rz,Bern:2008ap,Drummond:2008aq} more manifest.  The translation from momentum space to twistor space proceeds as follows.  Given a planar Feynman integral such as the one in fig.~\ref{fig:traintrack-integrals}, we introduce dual coordinates \(x_{\ell_i}\) for each loop and \(x_i\) for each external region.  Under the twistor correspondence, each of these dual points corresponds to a projective line \(\mathbb{P}^1\) inside a \(\mathbb{P}^3\) space.  This \(\mathbb{P}^3\) is called momentum twistor space.  Under this dictionary, the action of the conformal group on the dual space with coordinates \(x\) becomes the familiar \(\operatorname{PSL}(4)\) action on \(\mathbb{P}^3\).

Two dual points are light-like separated if their corresponding lines in momentum twistor space intersect.  This simple geometric fact, which is manifestly invariant under \(\operatorname{PSL}(4)\) transformations, will be central to our discussions below.  Indeed, the leading singularity locus is obtained by imposing a number of light-like conditions between the dual points.  Using the momentum twistor constructions these constraints yield a configuration of intersecting lines, which is much easier to describe than the set of quadratic equations which one has to solve in momentum space or dual space.

Another advantage of the momentum twistor description is that it automatically picks for us a compactification and complexification of the dual space which is compatible with the dual conformal symmetry.  The complexification is essential as well since all the varieties we will describe below are complex varieties.

Our analysis is similar in spirit to the analysis done by Hodges~\cite{Hodges:2010kq} for the one-loop box integral.  The one-loop box example is however much simpler, since its leading singularity locus is a set of two points.

In this paper we obtain the following results.  We describe the leading singularity locus of the elliptic double box as an intersection of two quadrics in \(\mathbb{P}^3\).  We compute the \(j\)-invariant of this intersection and compare with the answer obtained in ref.~\cite{Bourjaily:2017bsb}.  Next, we analyze the three-loop case and we identify the leading singularity locus with a K3 surface.  The K3 surface is described as a branched surface over the union of two genus-one curves in \(\mathbb{P}^1 \times \mathbb{P}^1\).  We compute its Euler characteristic and the number of moduli.  Then, we analyze the leading singularity locus in the four-loop case.  We obtain a Calabi-Yau three-fold which can be realized as a complete intersection.  We analyze its topology using the methods of Batyrev and Borisov.  Finally we end with short discussions of the higher-loop cases and of the supersymmetrization.

\section{Two loops: the elliptic double box}
\label{sec:double-box}

\subsection{Construction}

We consider the two-loop traintrack diagram, i.e.\ the two-loop version of the class of diagrams depicted in fig.~\ref{fig:traintrack-integrals}.
Its leading singularity is determined as follows.  There are three dual points \(x_1\), \(x_2\), \(x_3\) corresponding to the left loop and three dual points \(x_4\), \(x_5\), \(x_6\) corresponding to the right loop.  The left loop internal dual point \(x_{\ell_1}\) has to be light-like separated from the three dual points \(x_1\), \(x_2\), \(x_3\).  The right loop internal dual point \(x_{\ell_2}\) has to be light-like separated from the three dual points \(x_4\), \(x_5\), \(x_6\).  Finally, the points \(x_{\ell_1}\) and \(x_{\ell_2}\) have to be light-like separated.

In momentum twistor space this can be described as follows.  To each dual point \(x_i\) we associate a line \(A_i \wedge B_i\) in momentum twistor space \(\mathbb{P}^3\).  Two dual points are light-like separated if their corresponding lines in \(\mathbb{P}^3\) intersect.  At first, we assume that all the lines corresponding to external dual points are skew (do not meet in \(\mathbb{P}^3\)).  When some of these lines intersect, the geometry simplifies.

Given three skew lines, there is a one-dimensional family of lines which intersect all of them.  This can be seen by using several fundamental results about quadrics in \(\mathbb{P}^3\).  The first fact is that three skew lines uniquely determine a non-singular quadric \(Q\).  The second fact is that a non-singular quadric \(Q\) in \(\mathbb{P}^3\) contains two families of lines where the lines in a given family are skew while two lines in different families always intersect.  Finally, through a given point passes a unique line from each family of lines.  Such families of lines on a quadric are called \emph{rulings}.

More concretely, given three skew lines \(A_i \wedge B_i\) for \(i = 1, 2, 3\), the quadric they determine can be written as
\begin{align}
  \label{eq:quadric-P3-defining-equation}
  Q(Z) = \langle Z A_1 B_1 A_3 \rangle \langle Z A_2 B_2 B_3 \rangle - \langle Z A_1 B_1 B_3 \rangle \langle Z A_2 B_2 A_3 \rangle.
\end{align}
Here \(Z\), \(A_i\) and \(B_i\) are points in \(\P{3}\) and \(\langle A B C D \rangle = \det(A, B, C, D)\) is the usual four-bracket of momentum twistors.
The three lines appear symmetrically, but this is not manifest in the formula above.
Using Pl\"ucker relations one can show that the symmetry holds.

\begin{figure}[ht]
  \centering
  \begin{minipage}[b]{0.48\textwidth}
    \centering
    \includegraphics[width=0.75\textwidth]{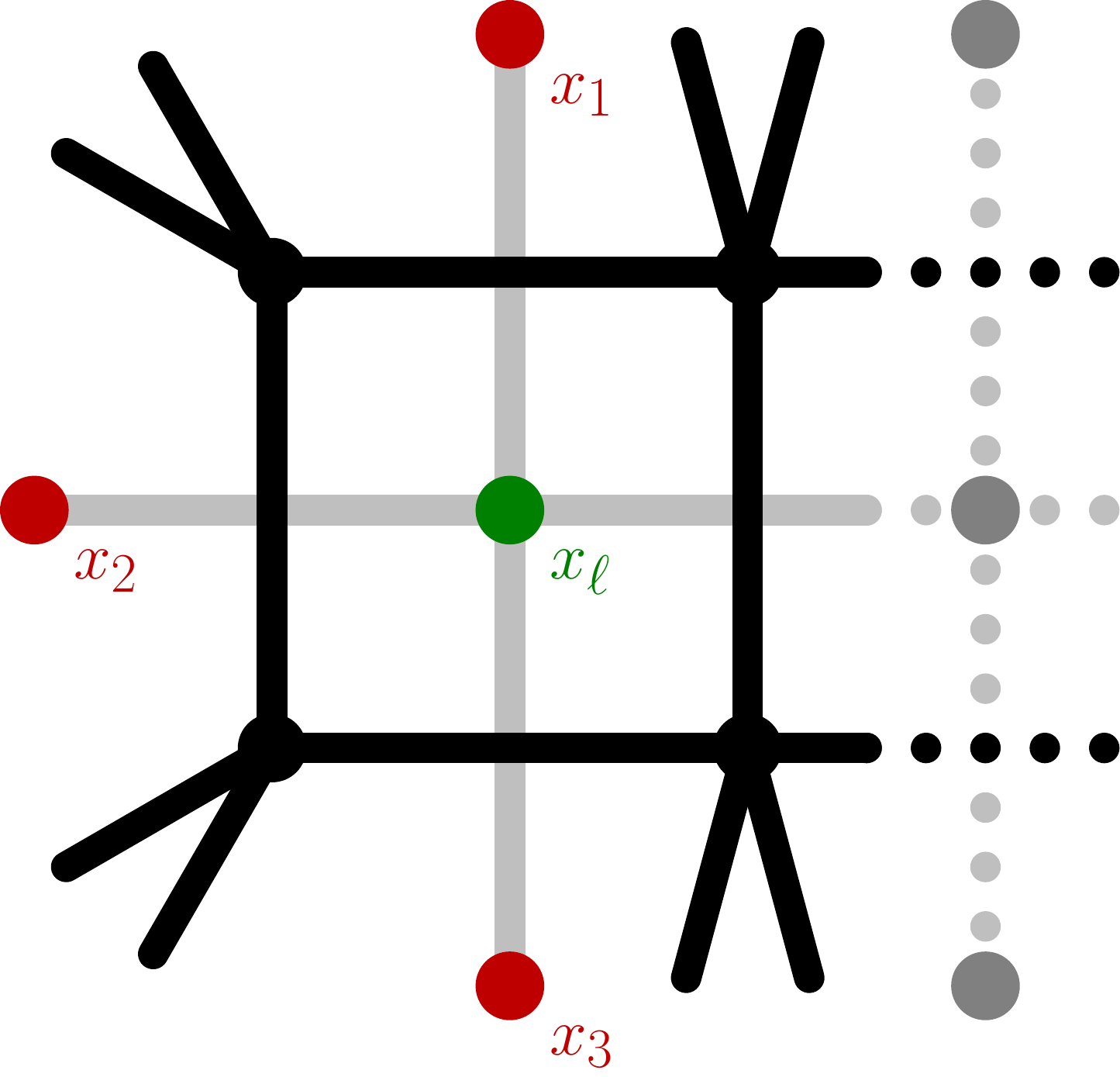}
  \end{minipage}
  \begin{minipage}[b]{0.48\textwidth}
    \centering
    \includegraphics[width=0.75\textwidth]{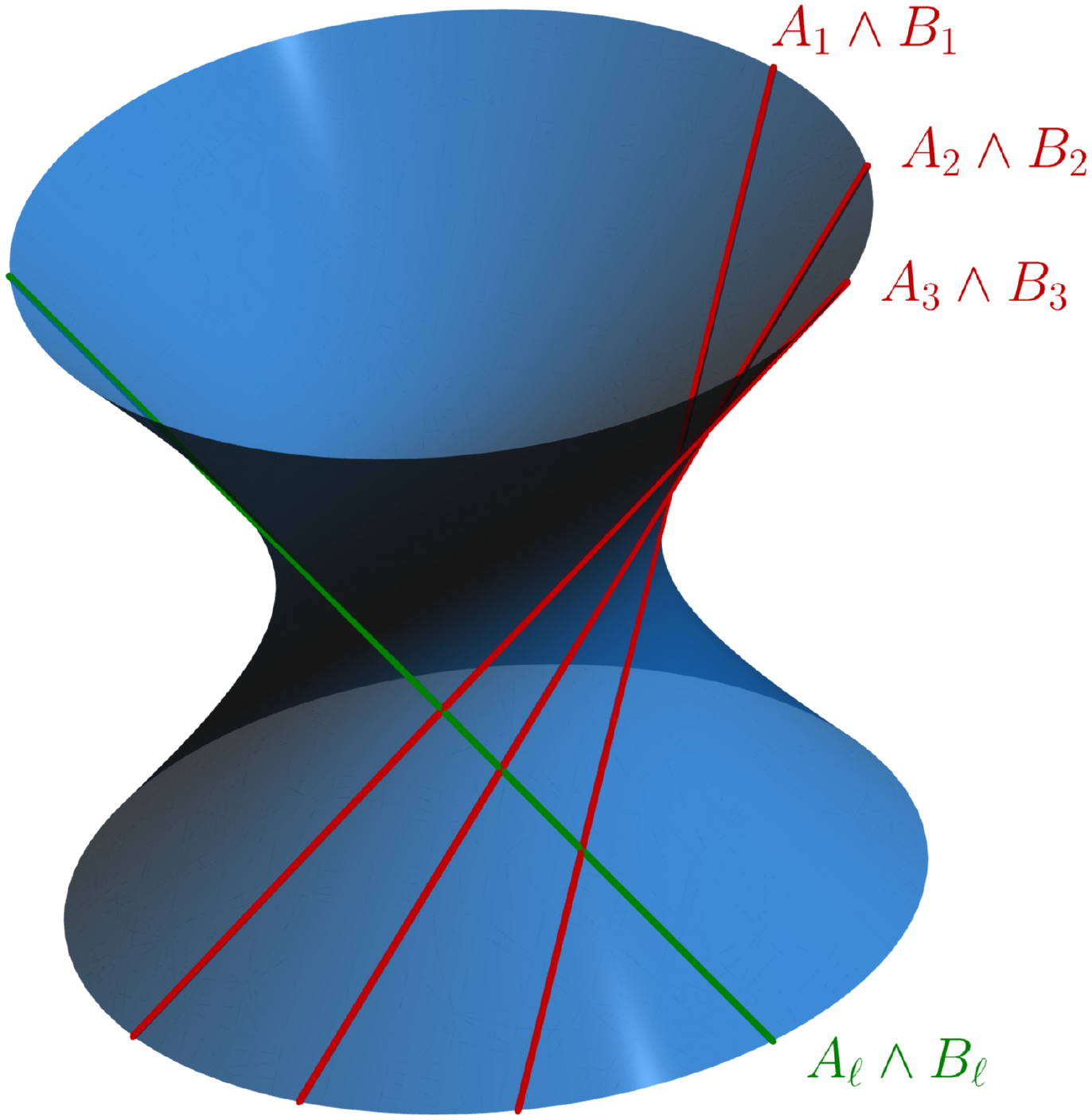}
  \end{minipage}
  \caption{Relationship between the endcap of the traintrack and the quadric.}
  \label{fig:traintrack-endcaps}
\end{figure}

Then, to the dual points \(x_1\), \(x_2\), \(x_3\) neighboring the left loop we can associate a quadric \(Q_L\) and to the points \(x_4\), \(x_5\), \(x_6\) neighboring the right loop we can associate a quadric \(Q_R\); cf.\ fig.~\ref{fig:traintrack-endcaps}.
Next, consider the intersection \(C \equivR Q_L \cap Q_R \subset \P{3}\) of these two quadrics, which is a curve.  To each point on \(C\) we can associate a line in \(Q_L\) which intersects all the three lines determining \(Q_L\).  This line corresponds to the interior dual point \(x_{\ell_1}\) of the left loop.  Similarly, through the same point of \(C\) we can construct a line which intersects all the lines in \(Q_R\) corresponding to the interior dual point \(x_{\ell_2}\).  The line in \(Q_L\) and the one in \(Q_R\) intersect in a point in \(C\) so their corresponding dual points are also light-like separated as required for the leading singularity.

\begin{figure}[t]
  \centering
  \includegraphics[width=0.9\textwidth]{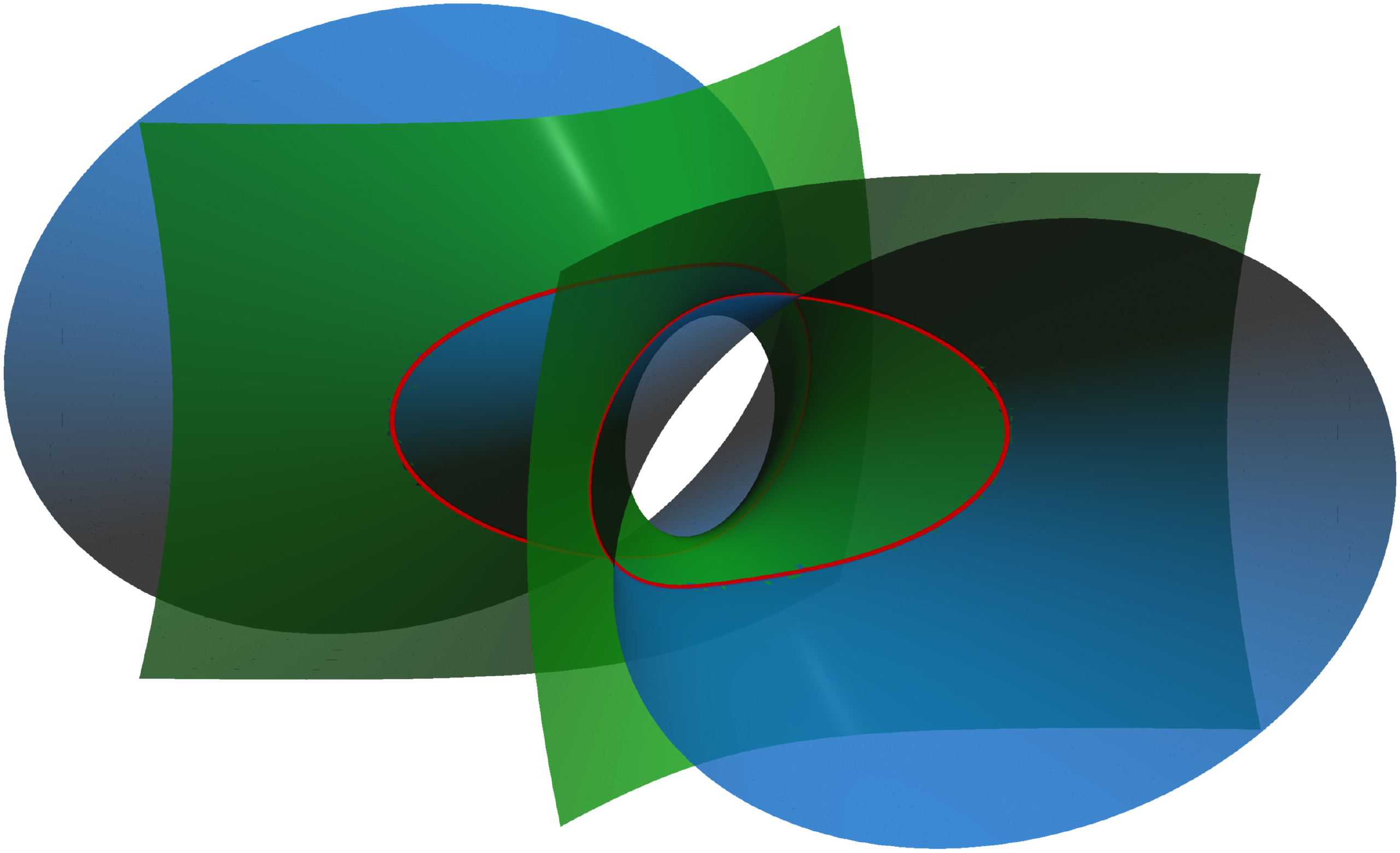}
  \caption{Two intersecting quadrics. Their intersection is the genus-one curve $C$ in the elliptic double box.}
  \label{fig:intersecting-quadrics}
\end{figure}

The intersection of two quadrics in \(\P{3}\) is a genus-one algebraic curve, see fig.~\ref{fig:intersecting-quadrics}.

We can connect this construction to the more familiar picture of a cubic curve in \(\mathbb{P}^2\) as follows:
Without loss of generality, we can take the point \([X_0 : X_1 : X_2 : X_3] = [0 : 0 : 0 : 1]\) to belong to both quadrics.  Then the equations for the two quadrics can be written as
\begin{equation}
  Q_L = X_3 L_L + M_L, \qquad
  Q_R = X_3 L_R + M_R,
\end{equation}
where \(L_L\) and \(L_R\) are of homogeneous of degree one and \(M_L\) and \(M_R\) are homogeneous of degree two in \(X_0\), \(X_1\) and \(X_2\).  When eliminating \(X_3\), we obtain \(L_L M_R - L_R M_L = 0\), which is a cubic in \(\mathbb{P}^2\).

\subsection{Analysis of the two-loop leading singularity locus}
\label{sec:analysis-elliptic-curve}

Having constructed a genus-one curve \(C\) as the intersection of two quadrics in \(\P{3}\), we now proceed to analyze its properties.

The holomorphic differential one-form on the curve can be found by taking Poincaré residues,
\begin{align}
  \label{eq:holomorphic_two_form}
  \omega_{C} = \Res_{Q_L} \Res_{Q_R} \frac{\omega_{\P{3}}}{Q_L Q_R}.
\end{align}
Here \(\omega_{\P{3}}\) is the \(\PSL(4)\)-invariant, weight-four holomorphic three-form on \(\P{3}\).
The quadrics \(Q_L\) and \(Q_R\) both have weight two so that the ratio \(\frac{\omega_{\P{3}}}{Q_L Q_R}\) is invariant under rescaling of the homogeneous coordinates of \(\P{3}\).
Then, we take two Poincaré residues which yields a one-form localized on \(C\).
This is in fact the unique holomorphic one-form on \(C\) so the curve \(C\) is indeed a genus-one curve.
A genus-one curve is characterized by only one modulus, which can be taken to be its \(j\)-invariant.

We can also see that there is only one modulus by counting parameters as follows:
There are six dual points, each with four coordinates.  From this, we need to subtract the dimension of the four-dimensional conformal group, which is \(15\).  In total we obtain \(6 \times 4 - 15 = 9\), assuming the conformal group acts effectively.  However, there are configurations of the three skew lines in the left quadric which generate the same quadric.  Indeed, consider a line inside \(Q_L\) which intersects all the lines which determine \(Q_L\).  We can displace any of these three lines along the chosen line without changing \(Q_L\).  Hence, there is a three-dimensional space of three skew lines which parametrize the same quadric \(Q_L\).  The same holds for \(Q_R\).  Moreover, the same curve \(C\) can be obtained by considering any two members of the so-called \emph{pencil of quadrics} generated by \(Q_L\) and \(Q_R\).\footnote{A pencil is a set of subvarieties, in this case quadrics, which are parametrized by a line~\cite{GriffithsHarris}.}  In other words, instead of using \(Q_L\) and \(Q_R\) we can use linear combinations of them, \(\lambda_L Q_L + \lambda_R Q_R\) and \(\mu_L Q_L + \mu_R Q_R\), where \([\lambda_L : \lambda_R]\) and \([\mu_L : \mu_R]\) are homogeneous coordinates on a projective line.
This amounts to two extra parameters which do not appear in the moduli of \(C\).  In the end, \(C\) has \(9 - 3 - 3 - 2 = 1\) moduli.

The pencil of quadrics \(\lambda_L Q_L + \lambda_R Q_R\) also allows us to compute the \(j\)-invariant of the curve \(C\).
As mentioned above, \(C\) is obtained as the intersection of any two members of the pencil.
We now think of each of the quadrics as a \(4 \times 4\) symmetric matrix of the coefficients in the defining equation~\eqref{eq:quadric-P3-defining-equation} and consider the determinant
\begin{align}
  \label{eq:det-pencil}
  \det(\lambda_L Q_L + \lambda_R Q_R).
\end{align}
This is a polynomial of degree four in the homogeneous coordinates \([\lambda_L : \lambda_R]\) of \(\P{1}\).
Hence, it vanishes at four points in \(\P{1}\) and we conclude that there are four singular members of the pencils.\footnote{Note that we assume that the quadrics \(Q_L\) and \(Q_R\) are in general position such that the four roots of~\eqref{eq:det-pencil} are distinct.  If they are not, then the intersection degenerates and the integral can be computed in terms of generalized polylogarithms.}
The cross-ratio of these four points is an invariant of the pencil.
More concretely, let us denote the four points where~\eqref{eq:det-pencil} vanishes by \(\lambda^i \equivR [\lambda_L^i : \lambda_R^i]\).
Then, we can form the cross-ratio \(z = \frac{\langle 1 2 \rangle \langle 3 4 \rangle}{\langle 1 3 \rangle \langle 2 4 \rangle}\), where \(\langle i j \rangle = \det(\lambda^i, \lambda^j)\), and the \(j\)-invariant 
\begin{align}
  \label{eq:j-invariant-double-box}
  j = 256 \frac{(z^2 - z + 1)^3}{z^2 (z - 1)^2}.
\end{align}

As pointed out above, the curve \(C\) is obtained as the intersection of \emph{any} two members of the pencil of quadrics \(\lambda_L Q_L + \lambda_R Q_R\).
Thus we can characterize isomorphism classes of \(C\) by completely characterizing the pencil.
The cross-ratio \(z\) formed above classifies the isomorphism classes of four ordered points on \(\P{1}\) up to projective equivalence.
The \(j\)-invariant formed in~\eqref{eq:j-invariant-double-box} has the correct symmetries for the corresponding elliptic curve:
In defining the cross-ratio \(z\), we have the freedom of permuting three of the points \(\lambda^i\) on \(\P{1}\) while keeping one fixed without changing \(C\).
This permutation acts on \(z\) by sending \(z \mapsto z' \in \{z, \frac{1}{z}, 1 - z, 1 - \frac{1}{z}, \frac{1}{1 - z}, 1 - \frac{1}{1 - z}\}\).
One can check that the \(j\)-invariant in~\eqref{eq:j-invariant-double-box} is invariant under this map.

In~\cite{Bourjaily:2017bsb}, the elliptic double box integral was analyzed using the method of direct integration.
Starting from a dual-conformally invariant expression, Feynman parameters were introduced and as many integrations as possible were performed in terms of multiple polylogarithms.
Eventually, the authors found a representation of the double box integral of the form
\begin{align}
  \int_{0}^{\infty} d \alpha \, \frac{H_3(\alpha)}{\sqrt{Q(\alpha)}}.
\end{align}
Here \(H_3\) is a combination of weight-three multiple polylogarithms and \(Q(\alpha)\) is a polynomial in \(\alpha\) of degree four with coefficients depending on conformal cross-ratios.
The equation \(y^2 = Q(\alpha)\) thus defines an elliptic curve.
We have checked that the \(j\)-invariant of this curve matches the \(j\)-invariant of the curve constructed directly in momentum twistor space above.
This is an encouraging result as it means that the geometry is not merely an artifact of the chosen parametrization but an intrinsic property of the leading singularity of the double box integral.

\section{Three and more loops}
\label{sec:more-loops}

\subsection{K3 surface}
\label{sec:k3-surface}

\subsubsection{Construction}

\begin{figure}[ht]
  \centering
  \includegraphics[width=0.9\textwidth]{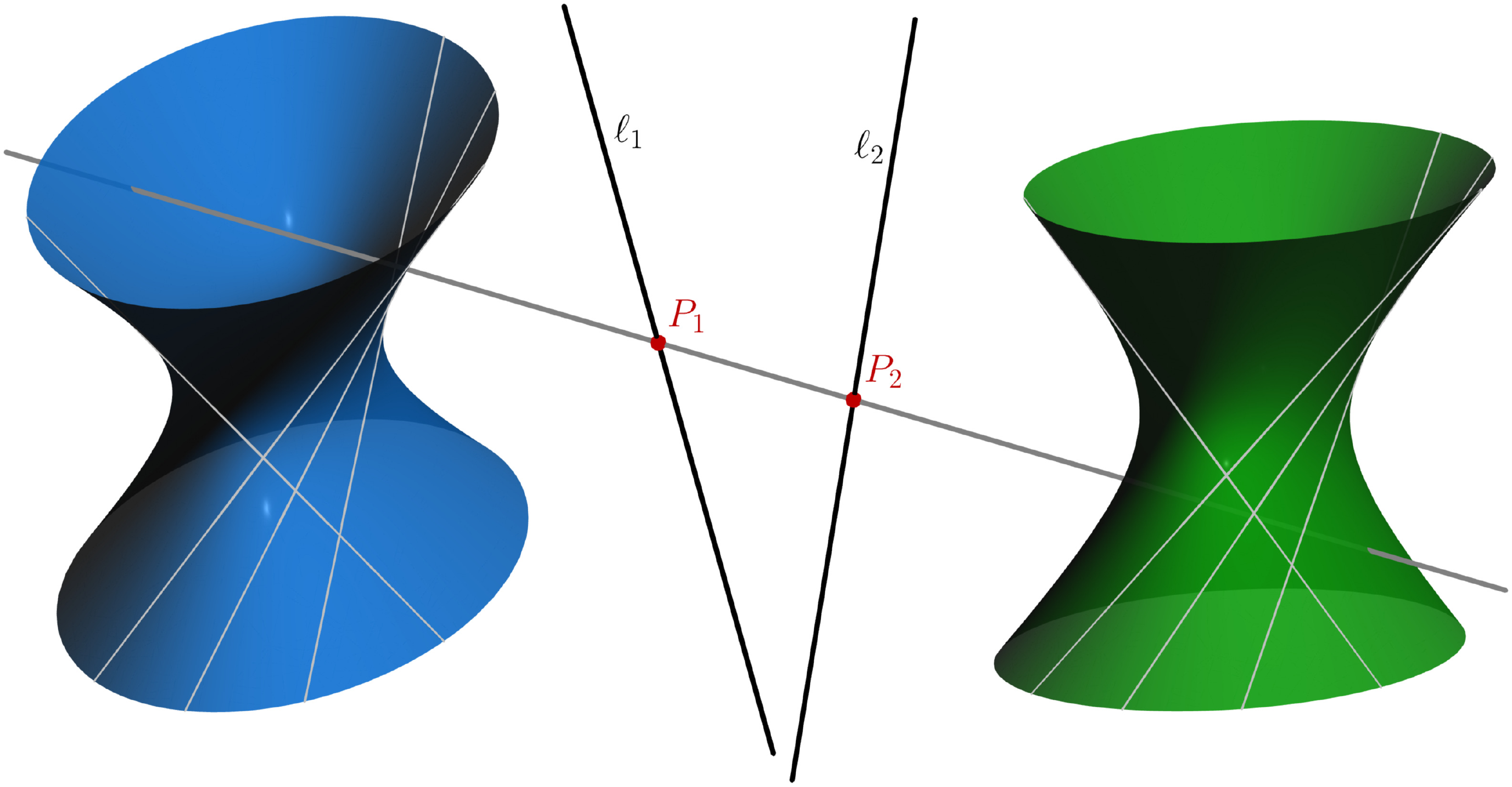}
  \caption{Quadrics and lines defining the K3 surface in the three-loop traintrack diagram.}
  \label{fig:quadrics-k3}
\end{figure}

The construction of a geometry for the three-loop traintrack integral is similar to the one for the two-loop case presented in section~\ref{sec:double-box}.
This time, however, we have two extra lines in momentum twistor space corresponding to the two additional external dual points.
The geometry in this case is given by two quadrics \(Q_L\) and \(Q_R\), constructed in the same way as at two loops, together with two lines \(\ell_1\) and \(\ell_2\).
Given points \(P_1 \in \ell_1\) and \(P_2 \in \ell_2\), we can construct a line \(P_1 \wedge P_2\) whose corresponding dual point is light-like separated from both dual points corresponding to \(\ell_1\) and \(\ell_2\).
The line \(P_1 \wedge P_2\) corresponds to the middle loop in the the three-loop traintrack integral.
The moduli space of these lines is \(\mathbb{P}^1 \times \mathbb{P}^1\) corresponding to the freedom in choosing \(P_1\) and \(P_2\).
We illustrate the construction in fig.~\ref{fig:quadrics-k3}.

The rest of the light-like constraints for the leading singularity can be imposed as follows.
By Bezout's theorem, the line \(P_1 \wedge P_2\) intersects the quadric \(Q_L\) in two points and the quadric \(Q_R\) in two points.\footnote{Bezout's theorem states that \(n\) hypersurfaces of degrees \(d_1, \dotsc, d_n\) in complex projective space \(\mathbb{P}^n\) intersect in \(d_1 \cdots d_n\) points, if the number of intersection points is finite~\cite{GriffithsHarris}.  In our case, the quadric has degree two, while a line can be seen as the intersection of two hyperplanes, each of degree one.  Hence, the intersection consists of two points.
}
Choosing one of these intersections in \(Q_L\) and one in \(Q_R\), we obtain a leading singularity configuration.
In total, there are four choices.
The total space of leading singularities is therefore a four-fold cover of \(\P{1} \times \P{1}\), branched over the curves where the line \(P_1 \wedge P_2\) is tangent to \(Q_L\) or \(Q_R\).

To find out where this branching arises, consider the points \(\alpha_1 P_1 + \alpha_2 P_2\) on the line \(P_1 \wedge P_2\).
The intersection with \(Q_L\) is given by the equation
\begin{align}
  \label{eq:intersection-points-quadric}
  \alpha_1^2 \, Q_L(P_1, P_1) + 2 \alpha_1 \alpha_2 \, Q_L(P_1, P_2) + \alpha_2^2 \, Q_L(P_2, P_2) = 0.
\end{align}
The line \(P_1 \wedge P_2\) is tangent to \(Q_L\) if this has a double root, i.e.\ when the discriminant with respect to \(\alpha_1\) or \(\alpha_2\) vanishes,
\begin{align}
  \Delta_L \equivR Q_L(P_1, P_2)^2 - Q_L(P_1, P_1) Q_L(P_2, P_2) = 0.
\end{align}
The polynomial \(\Delta_L\) is homogeneous of bidegree \((2, 2)\) in the coordinates of \(\P{1} \times \P{1}\) that parametrize the points \(P_1 \in \ell_1\) and \(P_2 \in \ell_2\).

A similar analysis can be done for the right quadric and we obtain another polynomial \(\Delta_R\) of bidegree \((2, 2)\).
The curves determined by \(\Delta_L\) and \(\Delta_R\) intersect in eight points.\footnote{To see why, consider first the intersection of such a genus-one curve with a line in \(\mathbb{P}^1 \times \mathbb{P}^1\) which sits at a point in the first or the second \(\mathbb{P}^1\).  It is easy to see that this intersection consists of two points.  Now, consider a degeneration of the biquadratic into four lines.  Two of the lines sit at a point in the first \(\mathbb{P}^1\) while the other two sit at a point in the second \(\mathbb{P}^1\).  Each one of them intersects the biquadratic in two points.  In total, there are eight intersection points.  As we deform from a singular curve consisting of four lines to a non-singular one, the number of intersections is conserved.  This type of argument is often used in Schubert problems (see ref.~\cite{ArkaniHamed:2010gh} for a detailed discussion).}
At these eight points, all the branches of the surface meet.
Over the remaining points of the curves determined by \(\Delta_L\) and \(\Delta_R\) there are only two branches, while over the remaining points of \(\P{1} \times \P{1}\) there are four branches.

The curves in \(\P{1} \times \P{1}\) defined by the vanishing locus of \(\Delta_L\) and \(\Delta_R\) are themselves genus-one curves as can be seen as follows.
If we choose coordinates \(x = [x_0 : x_1]\) and \(y = [y_0 : y_1]\) on \(\P{1} \times \P{1}\), then we can write the equation for a biquadratic as
\begin{align}
  \label{eq:general-biquadratic-P1P1}
  \Delta(x, y) = \sum_{a,b,a',b' = 0}^{1} A_{a b, a' b'} \, x_a x_b \, y_{a'} y_{b'},
\end{align}
where \(A\) is symmetric in the first and second pair of indices independently and thus has 9 independent components.
We now embed \(\P{1} \times \P{1}\) into \(\P{3}\) using the Segre map.
Concretely, we identify the homogeneous coordinates \([z_0 : z_1 : z_2 : z_3]\) on \(\P{3}\) with the coordinates on \(\P{1} \times \P{1}\) as
\begin{align}
  z_0 = x_0 y_0, \;
  z_1 = x_0 y_1, \;
  z_2 = x_1 y_0, \;
  z_3 = x_1 y_1.
\end{align}
The image of \(\P{1} \times \P{1}\) is then a quadric in \(\P{3}\) given by \(z_0 z_3 - z_1 z_2 = 0\).
The biquadratic~\eqref{eq:general-biquadratic-P1P1} becomes
\begin{align}
  \Delta(z) = \sum_{i,j = 0}^{3} \tilde{A}_{ij} \, z_{i} z_{j},
\end{align}
where \(\tilde{A}\) is a \(4 \times 4\) symmetric matrix that depends on the original coefficients \(A_{a b, a' b'}\).
This defines another quadric in \(\P{3}\).
The intersection of these two quadrics is a genus-one curve with only one modulus, as we have discussed before.

\subsubsection{Analysis}

The holomorphic two-form on the surface is
\begin{align}
  \label{eq:omega-K3}
  \omega_{K3} = \frac{\omega_{\P{1}} \, \omega_{\P{1}}}{\sqrt{\Delta_L} \sqrt{\Delta_R}}.
\end{align}
Notice that this ratio has the right homogeneity in \(\P{1} \times \P{1}\):
The first \(\omega_{\P{1}}\) has bidegree \((2, 0\)) while the second one has bidegree \((0, 2)\).
The polynomials \(\Delta_L\) and \(\Delta_R\) both have bidegree \((2, 2)\) so that~\eqref{eq:omega-K3} has homogeneity zero as required.

An analogous construction can be done for the simpler case of a genus-one curve in \(\P{2}\) as a two-fold branched cover over four points in \(\P{1}\).  In that case, we can define a polynomial \(P\) whose roots are the four points and the holomorphic form is \(\frac {\omega_{\mathbb{P}^2}}{\sqrt{P}}\).

\paragraph{Euler characteristic}

It is well-known that the Euler characteristic \(\chi\) of a K3 surface is \(24\), but we can directly compute this from the construction in momentum twistor space.
To do so, we will use the basic fact that \(\chi\) is additive under surgery.

According to the branching described above, the K3 surface \(S\) has only one branch on the points \(\P{1} \times \P{1}\) where the two curves \(\Delta_L\) and \(\Delta_R\) meet, i.e.\ for the points in \(\Delta_L \cap \Delta_R\).
For the points that lie on either of the two curves, i.e.\ for \(\Delta_L \cup \Delta_R \setminus \Delta_L \cap \Delta_R\), there are two branches.
In the complement of the two curves, i.e.\ in \(\P{1} \times \P{1} \setminus \Delta_L \cup \Delta_R\), there are four branches.
It follows that
\begin{align}
  \begin{split}
    \chi(S)
    &= 4 \left[\chi(\P{1} \times \P{1}) - \chi(\Delta_L \cup \Delta_R)\right]
    + 2 \left[\chi(\Delta_L \cup \Delta_R) - \chi(\Delta_L \cap \Delta_R)\right] \\
    &\quad + \chi(\Delta_L \cap \Delta_R) \\
    &= 4 \chi(\P{1} \times \P{1}) - 2 \chi(\Delta_L \cup \Delta_R) - \chi(\Delta_L \cap \Delta_R).
  \end{split}
\end{align}
Next, we use the fact that \(\chi(\P{1} \times \P{1}) = \chi(\P{1})^2\), \(\chi(\P{1}) = 2\) and \(\chi(\Delta_L \cup \Delta_R) = \chi(\Delta_L) + \chi(\Delta_R) - \chi(\Delta_L \cap \Delta_R)\).
The Euler characteristic of a point is one and the intersection \(\Delta_L \cap \Delta_R\) consists of eight points, thus we get \(\chi(\Delta_L \cap \Delta_R) = 8\).
Moreover, \(\Delta_L\) and \(\Delta_R\) are genus-one curves, thus \(\chi(\Delta_L) = \chi(\Delta_R) = 0\).
Finally, we get
\begin{align}
  \chi(S) = 4 \times 2 \times 2 - 2 \times (-8) - 8 = 24.
\end{align}
This is the expected number for a K3 surface which has Betti numbers \(b_0 = 1\), \(b_2 = 22\) and \(b_4 = 1\) with the odd Betti numbers vanishing.

\paragraph{Counting the number of moduli}

We would now like to count the number of moduli of these K3 surfaces.  This amounts to a counting of degrees of freedom of two genus-one curves in \(\P{1} \times \P{1}\), intersecting in eight points.
On top of that, there are moduli that roughly speaking describe the position of the quadrics corresponding to the endcaps of the traintrack integrals.

Before solving the first problem, recall the more familiar case of two cubic curves in the projective plane \(\P{2}\).  A cubic curve in the projective plane is a non-zero linear combination of ten monomials.  Hence, the set of cubic curves forms a \(\P{9}\).  The condition that a point belongs to a cubic curve imposes a linear condition in \(\P{9}\).  Given nine points in \emph{general} position, there is a single cubic curve which contains all of them.
The condition that the nine points be generic is essential here.  In fact, consider two cubics in the projective plane.  By Bezout's theorem, they intersect in nine points.  In this case, these nine points can not be generic since they do not uniquely determine a cubic curve.  In fact, they determine a pencil of cubics.

The theorem of Cayley-Bacharach states that if two plane cubics intersect in nine points, then any other cubic which passes through eight of them automatically passes through the ninth~\cite{GriffithsHarris}.\footnote{The Cayley-Bacharach theorem is essential in proving the associativity of the group law on a genus-one curve.}

Let us now return to genus-one curves in \(\mathbb{P}^1 \times \mathbb{P}^1\).  A biquadratic curve in \(\mathbb{P}^1 \times \mathbb{P}^1\) is a linear combination of nine monomials of bidegree \((2, 2)\).  Hence, these curves form a \(\mathbb{P}^8\).  As before, the condition that a point belongs to such a curve is a linear condition in \(\mathbb{P}^8\).  Hence, eight points in \emph{general} position uniquely determine a genus-one curve in \(\P{1} \times \P{1}\).

Next, consider two such biquadratic curves.  They intersect in eight points.
If the equations of the two biquadratics in homogeneous coordinates \(x = [x_0 : x_1]\) and \(y = [y_0 : y_1]\) of \(\P{1} \times \P{1}\) are
\begin{align}
  \label{eq:biquadric-eq-1}
  \Delta_{0 0}(y) x_0^2 + 2 \Delta_{0 1}(y) x_0 x_1 + \Delta_{1 1}(y) x_1^2 = 0, \\
  \label{eq:biquadric-eq-2}
  \Delta_{0 0}'(y) x_0^2 + 2 \Delta_{0 1}'(y) x_0 x_1 + \Delta_{1 1}'(y) x_1^2 = 0,
\end{align}
then the intersection points have \(y\) coordinates satisfying
\begin{equation}
  (\Delta_{0 0}' \Delta_{1 1} - \Delta_{0 0} \Delta_{1 1}')^2 +
  4 (\Delta_{0 0}' \Delta_{0 1} - \Delta_{0 0} \Delta_{0 1}')
  (\Delta_{1 1}' \Delta_{0 1} - \Delta_{0 1}' \Delta_{1 1}) = 0.
\end{equation}
Here \(\Delta_{ij}\) and \(\Delta_{ij}'\) are quadratic in \(y\) such that this is a degree-eight polynomial and that generically there are eight such intersection points.
For each of these values of \(y\) the corresponding value of \(x \in \mathbb{P}^1\) is given by
\begin{equation}
  2 (\Delta_{0 0}' \Delta_{0 1} - \Delta_{0 0} \Delta_{0 1}') x_0 + (\Delta_{0 0}' \Delta_{1 1} - \Delta_{0 0} \Delta_{1 1}') x_1 = 0.
\end{equation}

These eight points can not be in general position, otherwise there would be a unique biquadratic curve containing them.  For this case, we have a variant of the Cayley-Bacharach theorem, stating that if two biquadratic curves meet in seven points then they meet in the eighth as well.

Returning to the problem of counting the moduli, we see that we have to specify seven points in \(\P{1} \times \P{1}\) which amounts to 14 parameters.
From this we have to subtract \(2 \times 3\) parameters due to \(\PSL(2)\) transformations on each \(\P{1}\).
Moreover, we need to pick two members of the pencil of quadrics \(\lambda_L Q_L + \lambda_R Q_R\) which adds two additional moduli.
It turns out that there is one more modulus corresponding to the relative position of the left and right quadric along the middle line through the points \(P_1\) and \(P_2\).
In total, the number of moduli is
\begin{align}
  14 - 2 \times 3 + 2 + 1 = 11.
\end{align}

There is another, more direct way to establish 11 as an upper bound for the number of moduli:
The K3 surface only depends on the left and right quadrics and the two lines \(\ell_1\) and \(\ell_2\).
In dual space we have \(8 \times 4 - 15 = 17\), where we subtracted \(15\) due to the action of the conformal group.
As discussed in section~\ref{sec:analysis-elliptic-curve}, we can move each of the three lines defining a quadric up and down along a line from the opposite ruling without changing the quadric.
Thus we can subtract \(2 \times 3 = 6\) coordinates.
In total we get \(8 \times 4 - 15 - 6 = 11\) moduli.

For algebraic K3 surfaces, the sum of the dimension of the moduli space and the generic Picard rank has to equal \(20\) (see ref.~\cite{Aspinwall:1996mn}).  Since we found a moduli space of dimension \(11\), then the generic Picard rank should be \(9\).  Below, we find the same answer by looking at Nikulin involutions.

In~\cite{Bourjaily:2018ycu}, the authors analyzed the three-loop traintrack integral using Feynman parameters and identified a K3 surface as a hypersurface in a certain weighted projective space.
For a generic hypersurface in this space they found an upper bound of 18 for the number of moduli which is compatible with the number that we found above.
In the case of the elliptic curve we were able to compare the momentum twistor construction to the one found in Feynman-parametric integration using the \(j\)-invariant of the curve and found that they give the same geometry.
For the K3 surfaces, a more thorough study of their characteristics is needed to conclude whether or not they are equal.

\paragraph{Automorphisms and Nikulin involutions}

To further characterize the K3 surface \(S\), we study its automorphisms, in particular those automorphisms that leave the holomorphic two-form on \(S\) invariant.
Such automorphisms are called symplectic.
If \(f\) is a symplectic automorphism of finite order \(n\) and \(f \neq \id\), then one can show that the set of fixed points \(\Fix(f) \subset S\) is non-empty and finite.
Moreover, the number of fixed points satisfies \(1 \leq \left|\Fix(f)\right| \leq 8\) and depends only on the order \(n\) of \(f\), see for example ref.~\cite{Huybrechts:2016uxh}.
Nikulin~\cite{zbMATH03674247} also showed that the order \(n\) can at most be eight, i.e.\ \(n \leq 8\), which means that only the combinations of \(n\) and \(\left|\Fix(f)\right|\) in table~\ref{tab:automorphisms-k3} are possible.

Symplectic automorphisms of order two are called \emph{Nikulin involutions} and the corresponding number of fixed points is eight.
Such involutions are realized in our K3 surface as follows.

Consider the left quadric \(Q_L\) and the line \(P_1 \wedge P_2\) transversal to \(\ell_1\) and \(\ell_2\), see also fig.~\ref{fig:quadrics-k3}.
\(P_1 \wedge P_2\) intersects \(Q_L\) in two points and exchanging these two points constitutes an involution of the left quadric.
Recall that the points of intersection are given by the two roots of~\eqref{eq:intersection-points-quadric}.
Since this in a quadratic equation, the difference between the two roots is \(\sqrt{\Delta_L}\).
Thus, exchanging the two points of intersection, sends \(\sqrt{\Delta_L}\) to \(-\sqrt{\Delta_L}\).
The fixed points of this involution of the left quadric are the points of \(Q_L\) at which \(P_1 \wedge P_2\) becomes tangent, i.e.\ the points described by the genus-one curve \(\Delta_L = 0\) in \(\P{1} \times \P{1}\).
Since the map we described so far changes the sign of \(\sqrt{\Delta_L}\), the holomorphic two-form~\eqref{eq:omega-K3} also changes sign and we only obtain a Nikulin involution of the K3 surface if we perform the same involution on the right quadric.
The fixed points are then the eight intersection points of the curves \(\Delta_L\) and \(\Delta_R\) in \(\P{1} \times \P{1}\).

An involution which is not symplectic is the exchange of the two \(\mathbb{P}^1\) corresponding to the lines \(\ell_1\) and \(\ell_2\).  Indeed, under this transformation the holomorphic two-form in eq.~\eqref{eq:omega-K3} picks up a sign.

The existence of automorphisms implies a lower bound for the Picard number \(\rho(S)\) of the K3 surface~\cite{Huybrechts:2016uxh}.
For a Nikulin involution, i.e.\ a symplectic automorphism of order two, the bound is \(\rho(S) \geq 9\) (see Appendix.~\ref{sec:automorphisms-of-k3-surfaces}).
Since the Picard number plus the dimension of the moduli space are equal to 20, this bound is consistent with the counting of the moduli above.
In fact in our case the bound is satisfied, i.e.\ \(\rho(S) = 9\); for this case a complete description of the Picard lattice of \(S\) can be found in ref.~\cite{zbMATH05189579}.

\subsection{Three-fold and beyond}
\label{sec:threefold}

In this section, we demonstrate how we can build a Calabi-Yau manifold embedded in a toric variety for the four- and higher-loop traintrack integrals.
It was shown by Batyrev that mirror families of Calabi-Yau manifolds can be constructed as anticanonical hypersurfaces in toric varieties and that their Hodge numbers can be computed combinatorially by counting points in an associated pair of reflexive polytopes~\cite{Batyrev:1994hm}.
This construction was generalized to complete intersection Calabi-Yau (CICY) manifolds by Batyrev and Borisov using the nef-partitions of a reflexive polytope pair~\cite{Batyrev:1994pg,1993alg.geom.10001B}.
The Hodge numbers in this case can be computed by means of a recursive generating function; an implementation of this function is available in \texttt{PALP}~\cite{Kreuzer:2002uu}.\footnote{Note that technically the generating function computes the \emph{stringy} Hodge numbers introduced in~\cite{1997alg.geom.11008B}.}

\subsubsection{Three-fold}

\begin{figure}[ht]
  \centering
  \includegraphics[width=\textwidth]{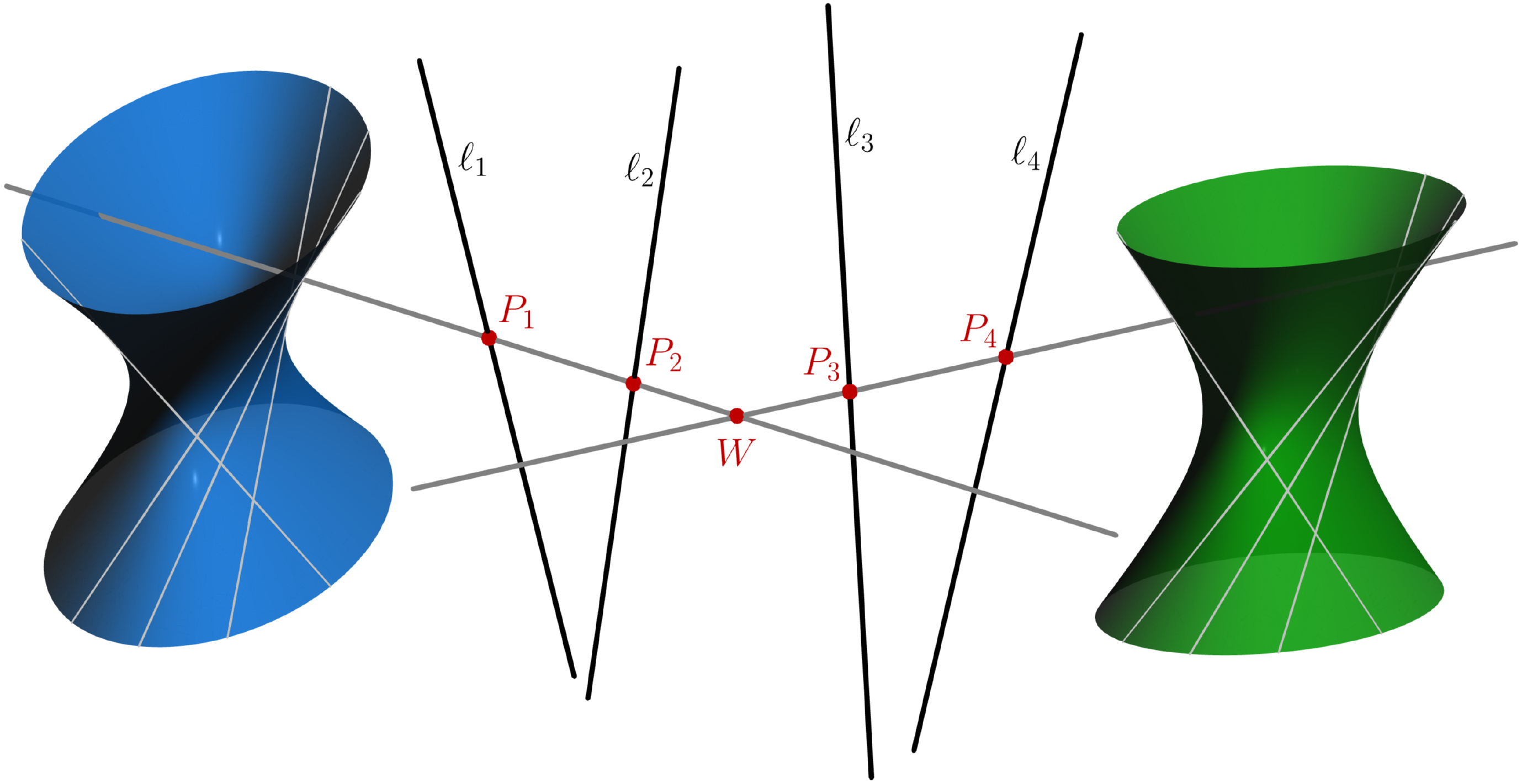}
  \caption{Quadrics and lines defining the CY three-fold in the four-loop traintrack diagram.}
  \label{fig:quadrics-cy3}
\end{figure}

The leading singularity configuration for the four-loop traintrack integral is depicted in fig.~\ref{fig:quadrics-cy3}.
Compared to the three-loop case discussed in section~\ref{sec:k3-surface}, we have two new lines, \(\ell_3\) and \(\ell_4\), corresponding to the two extra external dual points.

Let us introduce coordinates \(([\alpha_1 : \alpha_2], [\beta_1 : \beta_2])\) for the \(\P{1} \times \P{1}\) corresponding to the lines \(\ell_1\) and \(\ell_2\) and similarly \(([\gamma_1 : \gamma_2], [\delta_1: \delta_2])\) for the lines \(\ell_3\) and \(\ell_4\).
Then the embedding space is a toric variety defined by the relations
\begin{align}
  \label{eq:threefold-toric-variety-1}
  \begin{split}
    (\alpha_1, \alpha_2, \beta_1, \beta_2, y_L) &\sim (t_1\, \alpha_1, t_1\, \alpha_2, \beta_1, \beta_2, t_1\, y_L),
    \\
    (\alpha_1, \alpha_2, \beta_1, \beta_2, y_L) &\sim (\alpha_1, \alpha_2, t_2\, \beta_1, t_2\, \beta_2, t_2\, y_L)
  \end{split}
\end{align}
for the left part of fig.~\ref{fig:quadrics-cy3} and
\begin{align}
  \label{eq:threefold-toric-variety-2}
  \begin{split}
    (\gamma_1, \gamma_2, \delta_1, \delta_2, y_R) &\sim (t_3\, \gamma_1, t_3\, \gamma_2, \delta_1, \delta_2, t_3\, y_R),
    \\
    (\gamma_1, \gamma_2, \delta_1, \delta_2, y_R) &\sim (\gamma_1, \gamma_2, t_4\, \delta_1, t_4\, \delta_2, t_4\, y_R)
  \end{split}
\end{align}
from the right part.
Here \(t_1, t_2, t_3, t_4 \in \mathbb{C} \setminus \{0\}\) and the role of \(y_L\) and \(y_R\) will be clarified momentarily.
Since we have ten coordinates and four relations, we are left with a six-dimensional space.

Following the same construction as for the three-loop (K3) case, we obtain two polynomials \(\Delta_L\) and \(\Delta_R\) of bidegree \((2, 2)\) in \(\P{1} \times \P{1}\) from the left and right outermost loop of the traintrack.
In the six-dimensional toric variety constructed above, the Calabi-Yau manifold is defined as a codimension-three subvariety by means of the constraints
\begin{align}
  \label{eq:constraints-threefold}
  y_L^2 = \Delta_L, \quad
  y_R^2 = \Delta_R, \quad
  \langle P_1 P_2 P_3 P_4 \rangle = 0.
\end{align}
The last condition forces the two transversals \(P_1 \wedge P_2\) and \(P_3 \wedge P_4\) to intersect, see also fig.~\ref{fig:quadrics-cy3}.

The toric variety defined by the relations~\eqref{eq:threefold-toric-variety-1} and~\eqref{eq:threefold-toric-variety-2} can be described by a polytope with ten vertices in a six-dimensional integer lattice.
Explicitly, the vertices are given by the columns of the matrix
\begin{align}
  \label{eq:threefold-polytope}
  \begin{pmatrix*}[r]
    1  & 0  & 0  & 1  &-1  & 0  & 0  & 0  & 0  & 0 \\
    0  & 1  & 0  & 1  &-1  & 0  & 0  & 0  & 0  & 0 \\
    0  & 0  & 1  &-1  & 0  & 0  & 0  & 0  & 0  & 0 \\
    0  & 0  & 0  & 0  & 0  & 1  & 0  & 0  & 1  &-1 \\
    0  & 0  & 0  & 0  & 0  & 0  & 1  & 0  & 1  &-1 \\
    0  & 0  & 0  & 0  & 0  & 0  & 0  & 1  &-1  & 0
  \end{pmatrix*}.
\end{align}

The Hodge numbers of a generic codimension-three subvariety in this space can be obtained by computing the nef-partitions of the polytope defined by~\eqref{eq:threefold-polytope}.
Using \texttt{PALP}~\cite{Kreuzer:2002uu}, in particular the component \texttt{nef.x}\footnote{Note that we had to set \texttt{VERT\textunderscore{}Nmax} to 96 in \texttt{Global.h} for the computation to succeed.}, we find that there are 22 nef partitions.
Out of these, we identify three that have defining equations with degrees compatible with the constraints~\eqref{eq:constraints-threefold}.
The Hodge numbers are \(h^{11} = 12\) and \(h^{12} = 28\) which gives a Euler characteristic of \(\chi = -32\).

\subsubsection{General case}

The construction used for the three-fold, i.e.\ the four-loop case of the traintracks, generalizes to higher loops.
For \(L \geq 4\), we build a toric embedding space as follows:
There are \(2 + 4 (L - 2)\) coordinates, \(2\) from \(y_L\) and \(y_R\) and \(2 \times 2 (L - 2)\) from the two external dual points added with each loop.
The number of relations between these coordinates is \(2 (L - 2)\);
thus the dimension of the embedding space is \(2 + 4 (L - 2) - 2 (L - 2) = 2 (L - 1)\).
In this space, we impose \(2\) quadratic constraints, namely \(y_L^2 = \Delta_L\) and \(y_R^2 = \Delta_R\), as well as \(L - 3\) multilinear constraints.
Thus, the Calabi-Yau manifold is obtained as a subvariety of codimension \(L - 1\) in a toric variety of dimension \(2 (L - 1)\).
Note that the dimension of the manifold is also \(L - 1\).

As above, we can describe the embedding space by a polytope with vertices in an integer lattice.
The dimension of this lattice equals the dimension of the embedding space, i.e.\ \(2 (L - 1)\), while the number of vertices is equal to the number of coordinates, \(2 + 4 (L - 2)\).
The vertices are given in the general case by the columns of a block-diagonal matrix
\begin{align}
  \begin{pmatrix}
    \mathbf{A} & \mathbf{0} & \mathbf{0} & \cdots & \mathbf{0} \\
    \mathbf{0} & \mathbf{A} & \mathbf{0} &        &            \\
    \mathbf{0} & \mathbf{0} & \mathbf{B} &        &            \\
    \vdots     &            &            & \ddots & \vdots     \\
    \mathbf{0} &            &            & \cdots & \mathbf{B}
  \end{pmatrix},
  \qquad
  \mathbf{A} =
  \begin{pmatrix*}[r]
    1 & 0 & 0 &  1 & -1 \\
    0 & 1 & 0 &  1 & -1 \\
    0 & 0 & 1 & -1 &  0
  \end{pmatrix*},
  \qquad
  \mathbf{B} =
  \begin{pmatrix*}[r]
    1 & -1
  \end{pmatrix*}.
\end{align}
Note that in the case of the threefold (i.e.\ \(L = 4\)) that was discussed above, \(\mathbf{B}\) does not appear and the matrix reduces to~\eqref{eq:threefold-polytope}.

We note that the codimension grows with the loop order and this makes the analysis of these varieties in terms of complete intersections more challenging.  One may hope for a more ``efficient'' description of these varieties, but it remains to be seen if this is possible in way which is compatible with supersymmetry, as described in sec.~\ref{sec:supersymmetrization}.

\section{Supersymmetrization}
\label{sec:supersymmetrization}

The constructions presented so far are manifestly dual-conformal invariant.  Indeed, this is one reason why it makes sense to use momentum twistors to describe their geometry.  However, we know that the scattering amplitudes in \(\mathcal{N} = 4\) are in fact dual \emph{super}-conformal invariant.  It is then natural to ask what becomes of the supersymmetry.

In order to describe the supersymmetrization, we will redo the previous analysis in such a way that the various incidence relations are described in terms of \(\operatorname{PSL}(4)\)-invariant delta functions.  The basic ingredient will be the delta function of two points on \(\mathbb{P}^3\), which we denote by \(\delta_{\mathbb{P}^3}^3(P_1; P_2)\), where \(P_1, P_2 \in \mathbb{P}^3\).

This quantity can be used to define \(\delta^2_{\mathbb{P}^3}(L; P)\), which has support when the point \(P\) lies on the line \(L\).  If the line \(P\) contains two points \(P_0\) and \(P_1\), then we have
\begin{equation}
  \delta_{\mathbb{P}^3}^2(L; P) = \int \omega_{\mathbb{P}^1}(\alpha) \delta_{\mathbb{P}^3}(\alpha_0 P_0 + \alpha_1 P_1; P).
\end{equation}
Similarly, we can define \(\delta_{\mathbb{P}^3}(L_1; L_2)\), which has support when the two lines \(L_1\) and \(L_2\) intersect.

To define a delta function with support on a quadric, we use the fact that the quadric is determined by three skew lines \(L_1\), \(L_2\) and \(L_3\).  The quadric is ruled by a family of lines which intersect \(L_1\), \(L_2\) and \(L_3\).  Moreover, through any point on the quadric passes one line in this ruling.  We can then describe the conditions that a point \(P\) belongs to the quadric \(Q\) determined by the skew lines \(L_1\), \(L_2\) and \(L_3\) by the following integral
\begin{equation}
  \label{eq:quadric_delta}
  \delta_{\mathbb{P}^3}(Q; P) = \int \mu_{\mathbb{P}^3}(L) \delta_{\mathbb{P}^3}(L; L_1) \delta_{\mathbb{P}^3}(L; L_2) \delta_{\mathbb{P}^3}(L; L_3) \delta_{\mathbb{P}^3}^2(L; P),
\end{equation}
where \(\mu_{\mathbb{P}^3}(L)\) is the integral over the space of lines in \(\mathbb{P}^3\).  This integral is four-dimensional so, after performing the integrals, we are left with a single constraint.  This is expected since a quadric is of codimension one in \(\mathbb{P}^3\).

To obtain the genus-one curve we simply take the product of the two delta functions corresponding to \(Q_L\) and \(Q_R\).  This is a distribution which has support on the intersection of the two quadrics \(Q_L \cap Q_R\).  We can also obtain the holomorphic top form, but instead of taking Poincar\'e residues, we proceed as follows.  We look for a one-form \(\omega_C\) such that
\begin{equation}
  \int_C \omega_C(Z) f(Z) = \int \omega_{\mathbb{P}^3}(Z) \delta_{\mathbb{P}^3}(Q_L; Z) \delta_{\mathbb{P}^3}(Q_R; Z) f(Z),
\end{equation}
for any meromorphic function \(f\) on \(\mathbb{P}^3\) whose poles lie outside \(Q_L \cap Q_R\).

This construction is rather unnatural when done in \(\mathbb{P}^3\), but its advantage lies in the fact that it can be pretty straightforwardly supersymmetrized to \(\mathbb{P}^{3\vert 4}\).  Indeed, in \(\mathbb{P}^{3\vert 4}\) we have a delta function \(\delta_{\mathbb{P}^{3\vert 4}}^{3 \vert 4}(\mathcal{Z}_1; \mathcal{Z}_2)\), and so on.  These supersymmetrizations were introduced in ref.~\cite{Mason:2010yk}.  For the superquadric we obtain \(\delta_{\mathbb{P}^{3\vert 4}}^{1\vert 8}(\mathcal{Q}, \mathcal{Z})\).  Pursuing the same strategy as in the \(\mathbb{P}^3\) case, we finally define \(\omega_C^{1\vert 12}\) using
\begin{equation}
  \int_C \omega_C^{1\vert 12}(Z) f(Z) = \int \omega_{\mathbb{P}^{3\vert 4}}(\mathcal{Z}) \delta_{\mathbb{P}^{3\vert 4}}(\mathcal{Q}_l; \mathcal{Z}) \delta_{\mathbb{P}^{3\vert 4}}(\mathcal{Q}_r; \mathcal{Z}) f(Z),
\end{equation}
where \(\mathcal{Z} = [Z_0 : Z_1 : Z_2 : Z_3 \,\vert\, \chi_1 : \chi_2 : \chi_3 : \chi_4]\) and \(\omega_{\mathbb{P}^{3\vert 4}}(\mathcal{Z}) = \omega_{\mathbb{P}^3}(Z) d \chi_1 d \chi_2 d \chi_3 d \chi_4\) is the \(\operatorname{PSL}(4\vert 4)\)-invariant form on \(\mathbb{P}^{3\vert 4}\).

This construction can be generalized to higher dimensions.

\section{Summary and Outlook}
\label{sec:outlook}

We have presented a few examples of Calabi-Yau varieties arising as the leading singularity loci of the class of traintrack integrals.

For the elliptic double box we have a pretty explicit understanding of the moduli space and how it relates to the external kinematics of the integral.  We believe this should be a useful ingredient in the computation of these integrals.

The moduli space of algebraic K3 surfaces has a global description as a double coset of an orthogonal group (see ref.~\cite{Aspinwall:1996mn}).  This moduli space should be somehow parametrized by the external kinematics, but this global description does not seem to arise naturally from the twistor representation of the kinematics.  So, while we have described the topology of these varieties in some detail, our description of their moduli space has not been as detailed as we would like.  One approach we have sketched is to use a parametrization where \(10\) moduli arise from an intersection of two genus-one curves in \(\mathbb{P}^1 \times \mathbb{P}^1\) and an extra modulus arises from the intersections of transversals to these \(\mathbb{P}^1\) with the two quadrics \(Q_L\) and \(Q_R\).  It remains to be seen if this parametrization will be useful for expressing the corresponding integral.

One slightly mysterious aspect remains in connection with Calabi-Yau varieties encountered in non-planar integrals.  The twistor methods are well-adapted for studying planar integrals.  How should non-planar integrals be described in this language?  It is not clear yet if the momentum twistor approach is a useful description for the leading singularity locus of these integrals.  We hope to report on this issue in future work.

We have also discussed supersymmetrization.  The approach to supersymmetrization we have sketched generalizes to other cases as well.  Clearly supersymmetry imposes some restriction on the geometry of these varieties and it would be interesting to understand this better.

\vspace{\fill}
\section*{Acknowledgments}

We are grateful to Jacob Bourjaily, Andrew McLeod, Matt von Hippel and Matthias Wilhelm for discussions, collaboration on related topics and comments on the draft of this paper.
This work was supported in part by the Danish Independent Research Fund under grant number DFF-4002-00037 (MV), the Danish National Research Foundation (DNRF91), the research grant 00015369 from Villum Fonden and a Starting Grant \mbox{(No.\ 757978)} from the European Research Council (CV, MV).

\newpage
\appendix

\section{Automorphisms of K3 surfaces}
\label{sec:automorphisms-of-k3-surfaces}

For an account of the automorphisms of K3 surfaces see for example ref.~\cite[Chapter 15]{Huybrechts:2016uxh}.
In the following we summarize some of the most important facts.

When studying the group of automorphisms \(\Aut(S)\) of a K3 surface \(S\), one distinguishes between symplectic and non-symplectic automorphisms.
An automorphism \(f: S \rightarrow S\) of a K3 surface \(S\) is symplectic if the induced action on \(H^0(S, \Omega_S^2)\) is the identity, i.e.\ if it leaves the holomorphic two-form on \(S\) invariant.
One can show that \(\Aut(S)\) is discrete and that the subgroup \(\Aut_s(S) \subset \Aut(S)\) of symplectic automorphisms is of finite index, at least for projective K3 surfaces.

One can moreover show the following result:
Let \(f \in \Aut_s(S)\) be of finite order \(n\) and \(f \neq \id\).
Then the set of fixed points \(\Fix(f)\) is non-empty and finite and
\begin{align}
  \left|\Fix(S)\right| = \frac{24}{n} \prod_{p | n} \left(1 + \frac{1}{p}\right)^{-1}.
\end{align}
Moreover the number of fixed point satisfies \(1 \leq \left|\Fix(f)\right| \leq 8\) and only depends on the order \(n\) of \(f\).

Nikulin also proved that for \(f \in \Aut_s(S)\), the order \(n\) of \(f\) satisfies \(n \leq 8\).
This means that only the combinations of \(n\) and \(\left|\Fix(S)\right|\) shown in table~\ref{tab:automorphisms-k3} can occur.
For each \(n\), one can also derive a lower bound for the Picard number \(\rho(S)\) which is also shown in table~\ref{tab:automorphisms-k3}.
One can see that the Picard number of K3 surfaces with automorphisms tends to be quite high.

\begin{table}[t]
  \centering
  \begin{tabular}{c c c c c c c c}
    \toprule
    Order \(n\) & 2 & 3 & 4 & 5 & 6 & 7 & 8 \\
    \midrule
    \(\left|\Fix(S)\right|\) & 8 & 6 & 4 & 4 & 2 & 3 & 2 \\
    \(\rho(S) \geq\) & 9 & 13 & 15 & 17 & 17 & 19 & 19 \\
    \bottomrule
  \end{tabular}
  \caption{Symplectic automorphism orders and number of fixed points for a complex K3 surface \(S\). Here \(\rho(S)\) is the Picard number of \(S\). Table from ref.~\cite{Huybrechts:2016uxh}.}
  \label{tab:automorphisms-k3}
\end{table}

Symplectic automorphisms of order two were studied by Nikulin~\cite{zbMATH03674247} and are called \emph{Nikulin involutions}.
According to table~\ref{tab:automorphisms-k3}, a Nikulin involution of a complex K3 surface has eight fixed points and Picard number \(\rho(S) \geq 9\).
A classification of all algebraic K3 surfaces with Picard number satisfying the lower bound, i.e.\ \(\rho(S) = 9\) can be found in ref.~\cite{zbMATH05189579}.

\newpage

\bibliographystyle{physics}
\bibliography{amplitude_refs}

\providecommand{\href}[2]{#2}\begingroup\raggedright\begin{thebibliography}{10}

\bibitem{belkale2003}
P.~{Belkale} and P.~{Brosnan}, ``{Matroids, motives, and a conjecture of
  Kontsevich.},'' \href{http://dx.doi.org/10.1215/S0012-7094-03-11615-4}{{\em
  {Duke Math. J.}} \textbf{116} (2003) no. 1, 147--188}.

\bibitem{Brown:2010bw}
F.~Brown and O.~Schnetz, ``{A K3 in $\phi^4$},''
\href{http://arxiv.org/abs/1006.4064}{{ arXiv:1006.4064 [math.AG]}}.

\bibitem{Cachazo:2008vp}
F.~Cachazo, ``{Sharpening The Leading Singularity},''
\href{http://arxiv.org/abs/0803.1988}{{ arXiv:0803.1988 [hep-th]}}.

\bibitem{Usyukina:1993ch}
N.~I. Usyukina and A.~I. Davydychev, ``{Exact Results for Three and Four Point
  Ladder Diagrams with an Arbitrary Number of Rungs},''
\href{http://dx.doi.org/10.1016/0370-2693(93)91118-7}{{\em Phys. Lett.}
  \textbf{B305} (1993)  136--143}.

\bibitem{CaronHuot:2012ab}
S.~Caron-Huot and K.~J. Larsen, ``{Uniqueness of Two-Loop Master Contours},''
  \href{http://dx.doi.org/10.1007/JHEP10(2012)026}{{\em JHEP} \textbf{1210}
  (2012)  026},
\href{http://arxiv.org/abs/1205.0801}{{ arXiv:1205.0801 [hep-ph]}}.

\bibitem{Bourjaily:2017bsb}
J.~L. Bourjaily, A.~J. McLeod, M.~Spradlin, M.~von Hippel, and M.~Wilhelm,
  ``{Elliptic Double-Box Integrals: Massless Scattering Amplitudes beyond
  Polylogarithms},''
  \href{http://dx.doi.org/10.1103/PhysRevLett.120.121603}{{\em Phys. Rev.
  Lett.} \textbf{120} (2018) no. 12, 121603},
\href{http://arxiv.org/abs/1712.02785}{{ arXiv:1712.02785 [hep-th]}}.

\bibitem{Brown:2009ta}
F.~C.~S. Brown, ``{On the Periods of Some Feynman Integrals},''
\href{http://arxiv.org/abs/0910.0114}{{ arXiv:0910.0114 [math.AG]}}.

\bibitem{Bourjaily:2018yfy}
J.~L. Bourjaily, A.~J. McLeod, M.~von Hippel, and M.~Wilhelm, ``{A (Bounded)
  Bestiary of Feynman Integral Calabi-Yau Geometries},''
  \href{http://dx.doi.org/10.1103/PhysRevLett.122.031601}{{\em Phys. Rev.
  Lett.} \textbf{122} (2019) no. 3, 031601},
\href{http://arxiv.org/abs/1810.07689}{{ arXiv:1810.07689 [hep-th]}}.

\bibitem{Besier:2019hqd}
M.~Besier, D.~Festi, M.~Harrison, and B.~Naskrecki, ``{Arithmetic and geometry
  of a K3 surface emerging from virtual corrections to Drell--Yan
  scattering},'' \href{http://arxiv.org/abs/1908.01079}{{ arXiv:1908.01079
  [math.AG]}}.

\bibitem{Festi:2018qip}
D.~Festi and D.~van Straten, ``{Bhabha Scattering and a special pencil of K3
  surfaces},'' \href{http://dx.doi.org/10.4310/CNTP.2019.v13.n2.a4}{{\em
  Commun. Num. Theor. Phys.} \textbf{13} (2019)  463--485},
  \href{http://arxiv.org/abs/1809.04970}{{ arXiv:1809.04970 [math.AG]}}.

\bibitem{Bourjaily:2019hmc}
J.~L. Bourjaily, A.~J. McLeod, C.~Vergu, M.~Volk, M.~Von~Hippel, and
  M.~Wilhelm, ``{Embedding Feynman Integral (Calabi-Yau) Geometries in Weighted
  Projective Space},''
\href{http://arxiv.org/abs/1910.01534}{{ arXiv:1910.01534 [hep-th]}}.

\bibitem{Klemm:2019dbm}
A.~Klemm, C.~Nega, and R.~Safari, ``{The $l$-loop Banana Amplitude from GKZ
  Systems and relative Calabi-Yau Periods},''
  \href{http://dx.doi.org/10.1007/JHEP04(2020)088}{{\em JHEP} \textbf{04}
  (2020)  088}, \href{http://arxiv.org/abs/1912.06201}{{ arXiv:1912.06201
  [hep-th]}}.

\bibitem{Bloch:2014qca}
S.~Bloch, M.~Kerr, and P.~Vanhove, ``{A Feynman Integral via Higher Normal
  Functions},'' \href{http://dx.doi.org/10.1112/S0010437X15007472}{{\em Compos.
  Math.} \textbf{151} (2015) no. 12, 2329--2375},
\href{http://arxiv.org/abs/1406.2664}{{ arXiv:1406.2664 [hep-th]}}.

\bibitem{Bloch:2016izu}
S.~Bloch, M.~Kerr, and P.~Vanhove, ``{Local Mirror Symmetry and the Sunset
  Feynman Integral},''
  \href{http://dx.doi.org/10.4310/ATMP.2017.v21.n6.a1}{{\em Adv. Theor. Math.
  Phys.} \textbf{21} (2017)  1373--1453},
\href{http://arxiv.org/abs/1601.08181}{{ arXiv:1601.08181 [hep-th]}}.

\bibitem{Bourjaily:2018ycu}
J.~L. Bourjaily, Y.-H. He, A.~J. Mcleod, M.~Von~Hippel, and M.~Wilhelm,
  ``{Traintracks Through Calabi-Yaus: Amplitudes Beyond Elliptic
  Polylogarithms},''
  \href{http://dx.doi.org/10.1103/PhysRevLett.121.071603}{{\em Phys. Rev.
  Lett.} \textbf{121} (2018) no. 7, 071603},
\href{http://arxiv.org/abs/1805.09326}{{ arXiv:1805.09326 [hep-th]}}.

\bibitem{Hodges:2009hk}
A.~Hodges, ``{Eliminating Spurious Poles from Gauge-Theoretic Amplitudes},''
  \href{http://dx.doi.org/10.1007/JHEP05(2013)135}{{\em JHEP} \textbf{1305}
  (2013)  135},
\href{http://arxiv.org/abs/0905.1473}{{ arXiv:0905.1473 [hep-th]}}.

\bibitem{Drummond:2006rz}
J.~Drummond, J.~Henn, V.~Smirnov, and E.~Sokatchev, ``{Magic Identities for
  Conformal Four-Point Integrals},''
  \href{http://dx.doi.org/10.1088/1126-6708/2007/01/064}{{\em JHEP}
  \textbf{0701} (2007)  064},
\href{http://arxiv.org/abs/hep-th/0607160}{{ arXiv:hep-th/0607160}}.

\bibitem{Bern:2008ap}
Z.~Bern {\em et al.}, ``{The Two-Loop Six-Gluon MHV Amplitude in Maximally
  Supersymmetric Yang-Mills Theory},''
  \href{http://dx.doi.org/10.1103/PhysRevD.78.045007}{{\em Phys. Rev.}
  \textbf{D78} (2008)  045007},
\href{http://arxiv.org/abs/0803.1465}{{ arXiv:0803.1465 [hep-th]}}.

\bibitem{Drummond:2008aq}
J.~M. Drummond, J.~Henn, G.~P. Korchemsky, and E.~Sokatchev, ``{Hexagon Wilson
  Loop = Six-Gluon MHV Amplitude},''
  \href{http://dx.doi.org/10.1016/j.nuclphysb.2009.02.015}{{\em Nucl. Phys.}
  \textbf{B815} (2009)  142--173},
\href{http://arxiv.org/abs/0803.1466}{{ arXiv:0803.1466 [hep-th]}}.

\bibitem{Hodges:2010kq}
A.~Hodges, ``{The Box Integrals in Momentum-Twistor Geometry},''
  \href{http://dx.doi.org/10.1007/JHEP08(2013)051}{{\em JHEP} \textbf{1308}
  (2013)  051},
\href{http://arxiv.org/abs/1004.3323}{{ arXiv:1004.3323 [hep-th]}}.

\bibitem{GriffithsHarris}
P.~Griffiths and J.~Harris, {\em {Principles of Algebraic Geometry}}.
\newblock Wiley Classics Library. John Wiley \& Sons Inc., New York, 1978.

\bibitem{ArkaniHamed:2010gh}
N.~Arkani-Hamed, J.~L. Bourjaily, F.~Cachazo, and J.~Trnka, ``{Local Integrals
  for Planar Scattering Amplitudes},''
  \href{http://dx.doi.org/10.1007/JHEP06(2012)125}{{\em JHEP} \textbf{1206}
  (2012)  125},
\href{http://arxiv.org/abs/1012.6032}{{ arXiv:1012.6032 [hep-th]}}.

\bibitem{Aspinwall:1996mn}
P.~S. Aspinwall, ``{K3 surfaces and string duality},'' in {\em {Theoretical
  Advanced Study Institute in Elementary Particle Physics (TASI 96): Fields,
  Strings, and Duality}}, pp.~421--540.
\newblock 11, 1996.
\newblock \href{http://arxiv.org/abs/hep-th/9611137}{{ arXiv:hep-th/9611137}}.

\bibitem{Huybrechts:2016uxh}
D.~Huybrechts, {\em {Lectures on K3 Surfaces}}.
\newblock Cambridge University Press, 9, 2016.

\bibitem{zbMATH03674247}
V.~V. {Nikulin}, ``{Finite automorphism groups of K\"ahlerian surfaces of type
  K3.},'' {\em {Tr. Mosk. Mat. O.-va}} \textbf{38} (1979)  75--137.

\bibitem{zbMATH05189579}
B.~{van Geemen} and A.~{Sarti}, ``{Nikulin involutions on \(K3\) surfaces.},''
  \href{http://dx.doi.org/10.1007/s00209-006-0047-6}{{\em {Math. Z.}}
  \textbf{255} (2007) no. 4, 731--753}.

\bibitem{Batyrev:1994hm}
V.~V. Batyrev, ``{Dual polyhedra and mirror symmetry for Calabi-Yau
  hypersurfaces in toric varieties},'' {\em J. Alg. Geom.} \textbf{3} (1994)
  493--545, \href{http://arxiv.org/abs/alg-geom/9310003}{{
  arXiv:alg-geom/9310003}}.

\bibitem{Batyrev:1994pg}
V.~V. Batyrev and L.~A. Borisov, ``{On Calabi-Yau complete intersections in
  toric varieties},'' \href{http://arxiv.org/abs/alg-geom/9412017}{{
  arXiv:alg-geom/9412017}}.

\bibitem{1993alg.geom.10001B}
L.~Borisov, ``{Towards the Mirror Symmetry for Calabi-Yau Complete
  intersections in Gorenstein Toric Fano Varieties},'' {\em arXiv e-prints}
  (1993)  , \href{http://arxiv.org/abs/alg-geom/9310001}{{
  arXiv:alg-geom/9310001 [math.AG]}}.

\bibitem{Kreuzer:2002uu}
M.~Kreuzer and H.~Skarke, ``{PALP: A Package for analyzing lattice polytopes
  with applications to toric geometry},''
  \href{http://dx.doi.org/10.1016/S0010-4655(03)00491-0}{{\em Comput. Phys.
  Commun.} \textbf{157} (2004)  87--106},
\href{http://arxiv.org/abs/math/0204356}{{ arXiv:math/0204356 [math.NA]}}.

\bibitem{1997alg.geom.11008B}
V.~V. Batyrev, ``{Stringy Hodge numbers of varieties with Gorenstein canonical
  singularities},'' {\em arXiv e-prints} (1997)  ,
  \href{http://arxiv.org/abs/alg-geom/9711008}{{ arXiv:alg-geom/9711008
  [math.AG]}}.

\bibitem{Mason:2010yk}
L.~Mason and D.~Skinner, ``{The Complete Planar $S$-Matrix of
  $\mathcal{N}\!=\!4$ SYM as a Wilson Loop in Twistor Space},''
  \href{http://dx.doi.org/10.1007/JHEP12(2010)018}{{\em JHEP} \textbf{12}
  (2010)  018},
\href{http://arxiv.org/abs/1009.2225}{{ arXiv:1009.2225 [hep-th]}}.

\end{thebibliography}\endgroup

\end{document}